\documentclass[aps,superscriptaddress,twocolumn,showpacs,preprintnumbers,amsmath,amssymb]{revtex4}

\usepackage{mathrsfs}
\usepackage{amsmath}
\usepackage{amssymb}
\usepackage{graphicx}
\usepackage{dcolumn}
\usepackage{bm}
\usepackage{subfigure}
\usepackage{color}
\usepackage{ifpdf}
\usepackage{epstopdf}
\usepackage{CJK}
\usepackage[
  pdfstartview=FitH,%
  bookmarks=true,%
  bookmarksopen=true,%
  breaklinks=true,%
  colorlinks=true,%
  linkcolor=blue,anchorcolor=blue,%
  citecolor=blue,filecolor=blue,%
  menucolor=blue,pagecolor=blue,%
  urlcolor=blue]{hyperref}
\usepackage{float}

\begin{document}

\title{Atomic momentum patterns with narrower interval}

\author{Baoguo Yang}
\affiliation{School of Electronics Engineering and Computer Science, Peking University, Beijing 100871, China}
\author{Shengjie Jin}
\affiliation{School of Electronics Engineering and Computer Science, Peking University, Beijing 100871, China}
\author{Xiangyu Dong}
\affiliation{School of Electronics Engineering and Computer Science, Peking University, Beijing 100871, China}
\author{Zhe Liu}
\affiliation{School of Electronics Engineering and Computer Science, Peking University, Beijing 100871, China}
\author{Lan Yin}
\affiliation{School of Physics, Peking University, Beijing 100871, China}
\author{Xiaoji Zhou}\email{xjzhou@pku.edu.cn}
\affiliation{School of Electronics Engineering and Computer Science, Peking University, Beijing 100871, China}
\date{\today}

\begin{abstract}
  We studied the atomic momentum distribution for a superposition of Bloch states spreading in the lowest band of an optical lattice after the action of the standing wave pulse. By designing the imposing pulse on this superposed state, an atomic momentum pattern appears with narrower interval between the adjacent peaks that can be far less than the double recoil momentum. The patterns with narrower interval come from the superposition of the action of the designed pulse on many Bloch states with quasi-momenta over the first Brillouin zone, where for each quasi-momentum there is an interference among several lowest bands. Our experimental result of narrow interval peaks is consistent with the theoretical simulation. The patterns of multi modes with different quasi-momenta are helpful for precise measurement and atomic manipulation.
\end{abstract}


\maketitle

\section{introduction}

Ultracold atomic gases or Bose-Einstein condensates (BECs) in the optical lattice (OL) provide a unique opportunity for quantum simulation of many-body systems and realization of quantum computation and high-precision atomic clock~\cite{Morsch,Scarola,Bloch,Liu1,Liu2,Liu3,Larson1,Larson2,Derevianko}. By now, most of these studies focus on the Bloch state with a certain single quasi-momentum, where the atomic momentum distribution is characterized by interference peaks with interval $2 \hbar k_L$ which is the double recoil momentum of the atoms in an OL, with $k_L$ the wave vector of the laser forming the OL. However, the atomic momentum distribution for a superposed state with quasi-momenta spreading in the first Brillouin zone (FBZ) of an OL is seldom studied.

The difficulties to study this superposed state lie in preparation of this state. In this paper we first load the atoms in the superposed Bloch state of S and D bands with zero quasi-momentum by nonadiabatic shortcut method within tens of microseconds. Then the BECs are maintained in the OL and a harmonic trap for as long as $30$ms. Because of the collisions between atoms, the atoms in D-band transfer to S-band gradually, and in S-band there appears the superposed state of Bloch states with quasi-momenta varying from $-\hbar k_L$ to $\hbar k_L$ in the FBZ.

After we experimentally prepare the initial state in which the momentum distribution is a Gaussian-like shape, we demonstrate the different patterns with the narrower interval by manipulation of the OL standing wave pulse. These peaks with interval less than the double recoil momentum come from the superposition of states with different quasi-momenta, which is useful for measurement with much higher contrast such as the atom interferometer~\cite{Cronin}.

This paper is organized as follows. In Sec.~\ref{sec:Preparation}, we give the method for preparing the superposition of Bloch states in S-band with quasi-momenta spreading in the FBZ. In Sec.~\ref{sec:Theory_manipulation}, a single Bloch state is compared with the superposed state of many Bloch states, and theory for the manipulation of the superposed state is provided. In Sec.~\ref{sec:Design}, we demonstrate the experimental and theoretical results of the momentum patterns with narrower interval. In Sec.~\ref{sec:Discussion_conclusion}, we show the method for further increasing contrast of the patterns and conclude the paper.

\section{Preparation of the initial state with superposition of Bloch states of different quasi-momenta}\label{sec:Preparation}

\begin{figure}
 \includegraphics[width=0.37\textwidth]{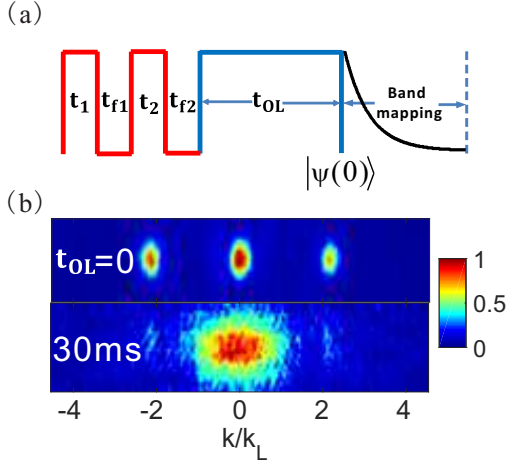}
  \caption{(a) The time sequences: after the first two pulses and the $30$ms holding time in the OL and the harmonic trap, the state becomes the superposition of the Bloch states in S-band with quasi-momenta spreading in FBZ, and is denoted by $\left| {\psi \left( 0 \right)} \right\rangle$. Then $1$ms band mapping is added. (b) The comparison of the experimental results between with and without ($t_{OL}=0$) the holding time.}\label{fig:time_sequences}
\end{figure}

There are adiabatic~\cite{PRA69.013603} and nonadiabatic~\cite{PRA68.051601,Liu} methods to load BECs into an OL as the Bloch state with a single quasi-momentum. To get the initial superposed state, we prepare BECs of $^{87}$Rb in a hybrid trap which is formed by the overlap of a single-beam optical dipole trap with the wave length $1064$nm and a quadruple magnetic trap. The condensate of about $1.5\times10^5$ atoms with the temperature $100$nK is achieved with the harmonic trapping frequencies $(\omega_x,\omega_y,\omega_z)=2\pi\times$(28Hz, 55Hz, 65Hz), respectively. Then using shortcut control method~\cite{Liu} we load the BECs into the state $\left( {\left| \text{S} \right\rangle  + \left| \text{D} \right\rangle } \right) / {\sqrt 2}$ with the quasi-momentum being zero, after evolution in two pulses consisting of OL pulses and intervals $({t_1,t_{f1},t_2,t_{f2}})$, as shown in Fig.~\ref{fig:time_sequences}(a). The lattice is produced by a standing wave created by two counter-propagating laser beams with lattice constant $a=\pi/k_L=426$nm along $x$ axis. We numerically calculate the pulses by modulating the parameters about the duration $t_{j}$ and the interval $t_{f_j}$ for the $j$th pulse. This loading process has above $99\%$ fidelity~\cite{Zhai} with $({t_1,t_{f1},t_2,t_{f2}})=(20.7,25.6,30.2,26.7)\mu$s for the OL depth $V_0$ being $10E_r$. Then we maintain BECs in the OL and the harmonic trap for as long as $30$ms. Because of the collisions between atoms, the atoms in D-band transfer to S-band gradually, and in S-band there appear atoms with quasi-momenta spreading in FBZ. It is crucial to choose an appropriate holding time of the OL, denoted as $t_{OL}$. When the time is short, the momentum components will oscillate versus the time~\cite{Xiong,Zhai}. However, for too long holding time, the system would be heating. By choosing $t_{OL}=30$ms, we obtain a superposed state of S-band Bloch states with quasi-momenta varying from $-\hbar k_L$ to $\hbar k_L$.

After retaining the superposed state $\frac{1}{\sqrt 2}\left( {\left| \text{S} \right\rangle  + \left| \text{D} \right\rangle } \right)$ in the OL and the harmonic trap for the time $t_{OL}$, followed by the $1$ms band mapping~\cite{Esslinger1,Spreeuw,Esslinger2,Bloch2} with the decay rate of about $100\mu$s as shown in Fig.~\ref{fig:time_sequences}(a), the absorption images after $25$ms time-of-flight (TOF) of BECs are given in Fig.~\ref{fig:time_sequences}(b) for $t_{OL}=0$ and $30$ms, which give the atomic density distribution in different bands. It is clear that for $30$ms holding time the atoms are almost spreading in the FBZ corresponding to the S-band, which is from the collisions during the holding time.

\section{Theory for manipulation of the superposed state}\label{sec:Theory_manipulation}
\begin{figure}
 \includegraphics[width=0.45\textwidth]{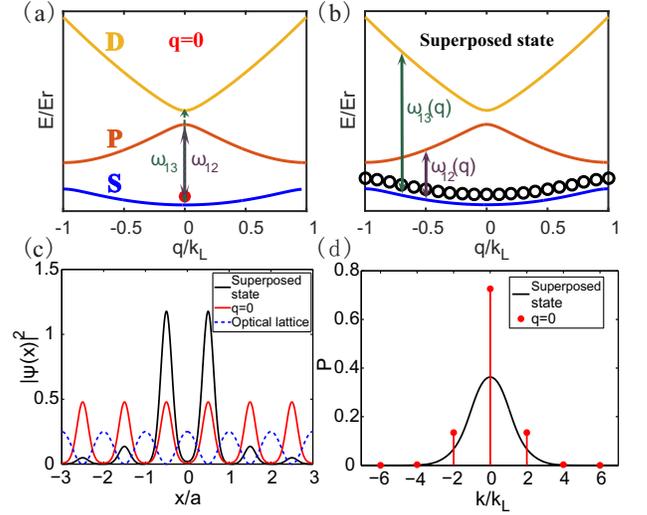}
  \caption{The schematic of the Bloch bands: (a) the Bloch state in S-band with $q=0$ is shown by the red dot; (b) the superposition of Bloch states of S-band in FBZ is shown with the black circles. (c) the wave functions modulus squared of the superposed state (the black line) and the Bloch state with $q=0$ (the red line) in coordinate space are shown, where the blue dashed line is the schematic of the OL. (d) the momentum distributions of the two states.}\label{fig:Initial_state}
\end{figure}

It is well known that the Bloch state has a periodical density distribution in coordinate space, and a discrete momentum component with interval $2 \hbar k_L$. However, what happens for the superposition of many Bloch states with different quasi-momenta? To find out the answer, we study the superposition of Bloch wave functions in S-band given by:
\begin{eqnarray}
 \left| {\psi \left( 0 \right)} \right\rangle  = \sum\limits_{q \in \left[ {-k_L,k_L} \right)} { \sqrt {f\left( q \right)\Delta q} \left| {{n_0},q} \right\rangle}.
\end{eqnarray}
$\hbar q$ is the quasi-momentum ($\hbar$ is the Plank constant), $n_0=1$ corresponds to the S-band, $\sqrt {f\left( q \right)\Delta q}$ is the proportional coefficient of the Bloch state $\left| {{n_0},q} \right\rangle$, and satisfies $\sum\limits_{q \in \left[ {-k_L,k_L} \right)} {f\left( q \right)\Delta q{\rm{ = }}1}$, and $\Delta q$ is the difference between two adjacent quasi-momenta. By setting $f\left( q \right)=1/\left( 2k_L \right)$ for convenience, we choose the initial state as:
\begin{eqnarray}
 \left| {\psi \left( 0 \right)} \right\rangle  = \frac{1}{\sqrt N } \sum\limits_{q \in \left[ {-k_L,k_L} \right)} {\left| {{n_0},q} \right\rangle},
\end{eqnarray}
with $N$ the number of the quasi-momenta in the summation among the FBZ. The Bloch state with $q=0$ is shown in Fig.~\ref{fig:Initial_state}(a) with the red dot and the superposed state $\left| {\psi \left( 0 \right)} \right\rangle$ is shown in Fig.~\ref{fig:Initial_state}(b) with black circles. Fig.~\ref{fig:Initial_state}(c) and (d) show the wave functions in coordinate and momentum space with the OL depth $10E_r$, respectively, where the black solid lines are the superposed state, and the red line (dots) corresponds to the Bloch state with $q=0$ as a contrast. The blue dashed line is the schematic of the OL. As shown in Fig.~\ref{fig:Initial_state}(c) and (d), the superposed state $\left| {\psi \left( 0 \right)} \right\rangle$ is a localized state in coordinate space, and a quasi-continuous distribution in momentum space when the number of quasi-momenta in FBZ $N$ tends to infinity.

Then we design the standing wave pulse imposing on this superposed state which is similar to the Kapitza-Dirac (KD) scattering of matter waves, which is a very powerful and versatile tool for the coherent splitting and mixing of momentum modes~\cite{Meystre,Pritchard}, and has been studied widely both in theory~\cite{Champenois,Chu,Clark,Phillips} and in experiments~\cite{APB69.303,PRL89.140401,PRL83.284,PRL83.5407,PRA80.043609,PRL83.3112,NJP11.013013,Xiong,PRL101.250401,PRL94.170403,Liu,Zhai,Yue,Hu,Niu}.
For the Bloch state with a single quasi-momentum, the effect of a sequence of OL standing wave pulses is superposition of different Bloch states with certain coefficients at the fixed quasi-momentum. Therefore, the resulting state has the peaks with $2 \hbar k_L$ interval in momentum space just as the the red dots in Fig.~\ref{fig:Initial_state}(d). However, it is entirely different for the superposed state of Bloch states with different quasi-momenta in the FBZ. We found that there appears an atomic momentum pattern with narrower interval between the adjacent peaks that can be chosen to be far less than the double recoil momentum.

To understand the action of the pulses on $\left| {\psi \left( 0 \right)} \right\rangle$, we start with the Hamiltonian:
\begin{eqnarray}
 \hat H =  - \frac{{{\hbar ^2}}}{{2M}}\frac{{{\partial ^2}}}{{\partial {x^2}}} + V\left( t \right){\cos ^2}\left( {{k_L}x} \right),
\end{eqnarray}
where $M$ is the atomic mass, $V\left( t \right)=0$ for $t \leqslant t_{11}$, and $V\left( t \right)=V_0$ for $t_{11} < t \leqslant t_{11}+t_{12}$. Wave function after the period of time $t_{11}$ is given by:
\begin{eqnarray}
 \left| {\psi \left( {t_{11}} \right)} \right\rangle  = \frac{1}{\sqrt N } \sum\limits_{q,{n_1}} {b\left( {q,{n_0},{n_1},{t_{11}}} \right)\left| {{n_1},q} \right\rangle },
\end{eqnarray}
where
\begin{eqnarray}
 b = \sum\limits_{{m_1}} {{c_{{n_0},q}}\left( {{m_1}} \right){c_{{n_1},q}}\left( {{m_1}} \right){e^{ - iE\left( {{m_1},q} \right){t_{11}}/\hbar }}}.
\end{eqnarray}
The coefficient ${{c_{n,q}}\left( m \right)}$ comes from the Bloch state:
\begin{eqnarray}
\left\langle x \right.\left| {n,q} \right\rangle  = {e^{iqx}}\sum\limits_{m = 0, \pm 1, \cdots } {{c_{n,q}}\left( m \right){e^{i2m{k_L}x}}},
\end{eqnarray}
and the kinetic energy is given by
\begin{eqnarray}
 E\left( {q,{m_1}} \right) = \frac{{\hbar ^2}{\left( {2{m_1}{k_L} + q} \right)^2}}{2M}.
\end{eqnarray}
After one single pulse, it has the form:
\begin{eqnarray}
 &&\left| {\psi \left( t_{11} + t_{12} \right)} \right\rangle  = \\
 &&\frac{1}{\sqrt N } \sum\limits_{q,{m_2}} {B\left( {q,{n_0},{m_2},{t_{11}},{t_{12}}} \right)\left| {\hbar \left( {2{m_2}{k_L} + q} \right)} \right\rangle }, \nonumber
\end{eqnarray}
with $\left| {\hbar \left( {2{m_2}{k_L} + q} \right)} \right\rangle$ the plane wave state, and
\begin{eqnarray}
 B = \sum\limits_{{n_1}} {b\left( {q,{n_0},{n_1},{t_{11}}} \right){e^{ - i{E_{{n_1},q}}{t_{12}}/\hbar }}{c_{{n_1},q}}\left( {{m_2}} \right)},
\end{eqnarray}
$E_{{n_1},q}$ is the eigen energy of the Bloch state with band $n_1$ and quasi-momentum $\hbar q$. The momentum distribution can be expanded as:
\begin{eqnarray}\label{eqn:P_2}
 &&P\left( {m_2},q \right) = \\
 &&\frac{1}{N}\sum\limits_{m,m',n,n'}{C \cos \left[ {W_{mm'}(q)}t_{11} + {\omega _{nn'}(q)}t_{12} \right]}, \nonumber
\end{eqnarray}
where
\begin{eqnarray}
 &&C \left( {m,m',n,n',m_2,q} \right) = \\
 &&{c_{1,q}}\left( m \right){c_{1,q}}\left( {m'} \right){c_{n,q}}\left( m \right){c_{n',q}}\left( {m'} \right){c_{n,q}}\left( {{m_2}} \right){c_{n',q}}\left( {{m_2}} \right), \nonumber
\end{eqnarray}
${W_{mm'}} = \left[ {E\left( {m,q} \right) - E\left( {m',q} \right)} \right]/\hbar $, and ${\omega _{nn'}} = \left( {{E_{n,q}} - {E_{n',q}}} \right)/\hbar$. $m=0, \pm 1, \pm 2, \cdots$ corresponds to different reciprocal momentum ($0, \pm 2 \hbar k_L, \pm 4 \hbar k_L, \cdots$), and $n=1, 2, 3, \cdots$ represents S, P, and D-band, etc. With definite evolution time $t_{11}$ and $t_{12}$, we can get figure of the probability versus the momentum.

By calculating the coefficient $C$, we learn that for terms with $W_{m,m'} \omega_{n,n'} \neq 0$, the coefficient $C$ is small enough, so we neglect these terms. Besides, the probability of momenta transformation between $-2 \hbar k_L$ and $2 \hbar k_L$ is far less than the probability between $2 \hbar k_L$ (or $-2 \hbar k_L$) and $0$, so we can ignore the terms including $W_{-1,1}$ and $W_{1,-1}$ in the summation of Eq.~(\ref{eqn:P_2}). Similarly, for the initial state beginning from S-band, the probability of transformation between P-band and D-band is far less than the probability between S-band and P-band (or D-band), and thus the terms with $\omega_{2,3}$ and $\omega_{3,2}$ can be neglected. Therefore, the Eq.~(\ref{eqn:P_2}) can be approximated as:
\begin{eqnarray}
 &&P\left( {0,q} \right) \approx {C_1} + {C_2}\cos \left( {{W_{1,0}}{t_{11}}} \right) + {C_3}\cos \left( {{W_{ - 1,0}}{t_{11}}} \right) \nonumber\\
 &&+ {C_4}\cos \left( {{\omega _{12}}{t_{12}}} \right) + {C_5}\cos \left( {{\omega _{13}}{t_{12}}} \right)
\end{eqnarray}
where the corresponding amplitudes $C_{i}\left( q \right)>0$ ($i=1, 2, \cdots, 5$) which comes from the numerical calculations. Therefore, at some given $q$, there will be peaks (valleys), as long as $W_{1,0}t_{11}$, $W_{-1,0}t_{11}$, $\omega_{12}t_{12}$ and $\omega_{13}t_{12}$ are all even (odd) times of $\pi$, from which the time $t_{11}$ and $t_{12}$ can be defined. By taking peaks as an example, we have $W_{1,0}t_{11}=2 l_1 \pi$, $W_{-1,0}t_{11}=2 l_2 \pi$, $\omega_{12}t_{12}=2 l_3 \pi$ and $\omega_{13}t_{12}=2 l_4 \pi$, so there are:
\begin{eqnarray}
 \frac{W_{1,0}}{W_{-1,0}} = \frac{{1 + q}}{{1 - q}} = \frac{l_1}{l_2},
\end{eqnarray}
and
\begin{eqnarray}
 \frac{\omega_{12}}{\omega_{13}} = \frac{l_3}{l_4},
\end{eqnarray}
with $l_{i}$ ($i=1, 2, 3, 4$) integers. For a tried quasi-momentum $q$, when $(1 + q)/(1 - q)$ or $\omega_{12}/\omega_{13}$ is irrational number, it can't be written as a fraction, and we can only choose an approximate value of $l_1/l_2$ or $l_3/l_4$, respectively. However, the numerical calculation can give an accurate results.

From Eq.~(\ref{eqn:P_2}), it is clear that the probability $P$ is the combination of disparate oscillations with different frequencies. The amplitudes and $W_{mm'}$ are all dependent on the momenta, and $\omega _{nn'}$ depends on the quasi-momentum. For different momenta, the probabilities reach maximum at different time intervals. Therefore, with some specific pulse, the probability distribution after the action of the pulse on the initial superposed state, can exhibit several peaks in the momentum space. By gradually increasing $t_{11}$ and $t_{12}$ from $0$ to $100\mu$s with a time step $1\mu$s, we can choose the appropriate $t_{11}$ and $t_{12}$, so that the momentum distribution has high contrast between peaks and the next valleys. The following pulses acting on the superposed state are all from the numerical calculations.

\begin{figure}
 \includegraphics[width=0.45\textwidth]{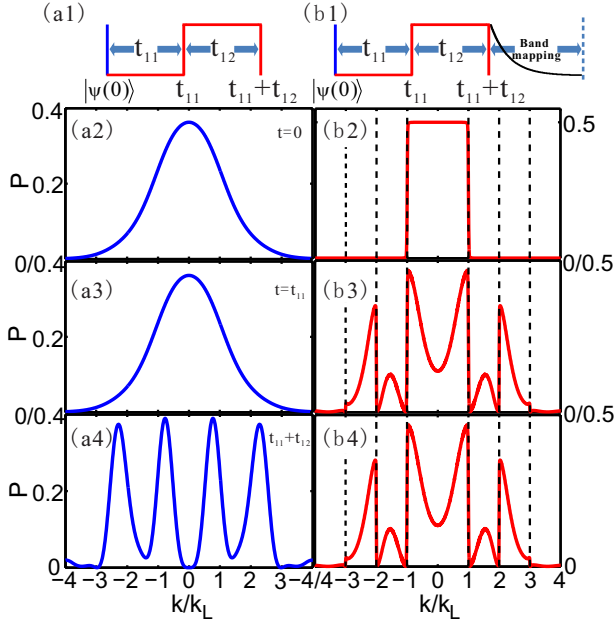}
  \caption{The single pulse acted on the superposed state $\left| {\psi \left( 0 \right)} \right\rangle$ without (a1) and with (b1) the band mapping. (a2-a4) show the momentum distributions at $t=0$, $t=t_{11}$ and $t=t_{11}+t_{12}$. (b2-b4) show the results of the band mapping on the basis of corresponding states in (a2-a4).
 }\label{fig:comparison}
\end{figure}

To visualize the evolution process, we give the momentum distributions without (Fig.~\ref{fig:comparison}(a2-a4)) and with band mapping (Fig.~\ref{fig:comparison}(b2-b4)). The time sequences are shown in Fig.~\ref{fig:comparison}(a1) and (b1). The band mapping reflects the population of the atoms in different bands. Fig.~\ref{fig:comparison}(a2) and (b2) are the results of $\left| {\psi \left( 0 \right)} \right\rangle$ at $t=0$. After a period of free evolution at $t_{11}=32\mu$s, the momentum distribution without band mapping is invariant (Fig.~\ref{fig:comparison}(a3)), while the proportion in different bands redistributes (Fig.~\ref{fig:comparison}(b3)). Then imposing the OL on BECs for $t_{12}=24\mu$s, the momentum distribution without band mapping shows four peaks between $-3\hbar k_L$ and $3\hbar k_L$ (Fig.~\ref{fig:comparison}(a4)). However, the proportion in different bands is invariable for the Bloch states being eigenstates of the Hamiltonian during the period of the time for imposing OL.

Similarly, after $l$ pulses, we have:
\begin{eqnarray}
 &&\left| {\psi \left( {t = \sum\limits_{i = 1}^l {\left( {{t_{i1}} + {t_{i2}}} \right)} } \right)} \right\rangle  = \\
 &&\frac{1}{\sqrt N } \sum\limits_{q,{m_{l + 1}}} {B \left( {q,{n_0},{m_{l + 1}},\{ t\} } \right)\left| {\hbar \left( {2{m_{l + 1}}{k_L} + q} \right)} \right\rangle }, \nonumber
\end{eqnarray}
where $\{ t\}  = {t_{11}},{t_{12}}, \cdots ,{t_{l1}},{t_{l2}}$, and
\begin{eqnarray}
 &&B \left( {q,{n_0},{m_{l + 1}},\{ t\} } \right) = \\
 &&\sum\limits_{{n_1}, \cdots ,{n_l}} {\prod\limits_{i = 1}^l {b\left( {q,{n_{i - 1}},{n_i},{t_{i1}}} \right){e^{ - i{E_{{n_i},q}}{t_{i2}}/\hbar }}} {c_{{n_l},q}}\left( {{m_{l + 1}}} \right)}, \nonumber
\end{eqnarray}
with
\begin{eqnarray}
 &&b\left( {q,{n_{i - 1}},{n_i},{t_{i1}}} \right) = \\
 &&\sum\limits_m {{c_{{n_{i - 1}},q}}\left( m \right){c_{{n_i},q}}\left( m \right){e^{ - iE\left( {m,q} \right){t_{i1}}/\hbar }}}. \nonumber
\end{eqnarray}
Therefore, the probability of momentum ${\hbar \left( {2{m_{l + 1}}{k_L} + q} \right)}$ after scattered by $l$ pulses is given by
\begin{eqnarray}
 &&P\left( {m_{l + 1},q} \right) = \\
 &&\frac{1}{N} {\left| {\sum\limits_{{n_1}, \cdots ,{n_l}} {\left[ {\prod\limits_{i = 1}^l {b {e^{ - i{E_{{n_i},q}}{t_{i2}}/\hbar }}} } \right]{c_{{n_l},q}}\left( {{m_{l + 1}}} \right)} } \right|^2}. \nonumber
\end{eqnarray}

\section{Experimental and theoretical demonstration for Patterns with narrower interval}\label{sec:Design}

\begin{figure}
 \includegraphics[width=0.45\textwidth]{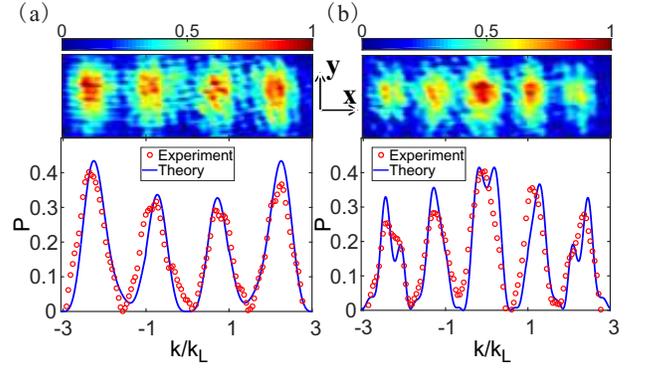}
  \caption{Patterns from the experimental measurements (the red circles) and the theoretical curves (the blue solid lines) with the different design, respectively, (a) four main peaks with pulse sequences $t_{11},t_{12}$ as ($32 \mu$s, $24 \mu$s) and (b) five main peaks with ($74 \mu$s, $47 \mu$s) between $\pm 3 \hbar k_L$ for OL depth $V_0=10E_r$. The color maps are the absorption images after a $25$ms TOF.} \label{fig:Theory_Experi1}
\end{figure}

In Fig.~\ref{fig:Theory_Experi1}, we give the different design for patterns of multi modes with various numbers of peaks under OL depth $10E_r$, where the color maps are the absorption images after the action of the pulse and the $25$ms TOF. To reduce the background noise, each result is the average over three absorption images with the same conditions, and the high frequency part of the images is removed. The red circles are the results of the average values along the y-axis among $40$ pixels of the absorption images. The blue solid lines are the numerical simulation, where we have fitted the quasi-momenta distribution $f\left( q \right)$ of the initial superposed state from the experimental result as shown in Fig.~\ref{fig:time_sequences}(b) with $t_{OL}=30$ms. In our measurement, the final atoms consist of condensed atoms and thermal atoms, while the thermal atoms without the quantum coherence do not attend the interference effect. Therefore, we have extracted the envelope of incoherent atoms, and only condensed atoms are left. In Fig.~\ref{fig:Theory_Experi1}(a), the pulse is given by $(t_{11},t_{12})=(32,24)\mu$s, and the pattern has four main peaks between $\pm 3 \hbar k_L$ with interval about $1.5 \hbar k_L$. Fig.~\ref{fig:Theory_Experi1}(b) gives the five main peaks with interval about $1.25 \hbar k_L$.

\begin{figure}
 \includegraphics[width=0.45\textwidth]{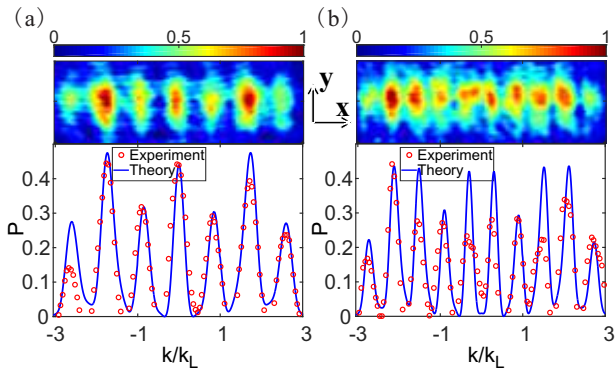}
  \caption{Patterns for the experimental measurements (the red circles) and the theoretical curves (the blue solid lines) with (a) seven main peaks with ($78 \mu$s, $20 \mu$s) and (b) ten main peaks with ($118 \mu$s, $19 \mu$s) between $\pm 3 \hbar k_L$ for OL depth $V_0=10E_r$. The color maps are the absorption images after a $25$ms TOF.}\label{fig:Theory_Experi2}
\end{figure}

Fig.~\ref{fig:Theory_Experi2}(a) and (b) show the patterns at $10E_r$ OL depth with seven main peaks ($0.87 \hbar k_L$ interval) and ten main peaks ($0.6 \hbar k_L$ interval), respectively. As the number of peaks increases, the average number of condensed atoms in each peak decreases, which results in that the influence of the background noise enhances, and the contrast reduces. Therefore we have averaged over six absorption images with the same conditions to get each experimental result, and removed the high frequency part of the absorption images. Similarly, we have removed the thermal atoms and remained the condensed part, in which the quantum coherence is maintained.

As shown in Fig.~\ref{fig:Theory_Experi1} and Fig.~\ref{fig:Theory_Experi2}, the agreement between theoretical simulations and experimental results is good, although the interatomic interaction is ignored, because the action time of the pulse is short, and effect of the interaction is very weak. The differences between the theory and the experiment mainly come from the heating effect of BECs and the inevitable experimental noise such as noise in intensity of the laser field and the inelastic scattering of photons. As the number of peaks increases and the interval between the nearest neighbor peaks decreases, the differences between the theory and the experiment increase. By improving the stability of the experimental system and getting colder atoms, there will be better consistency between the experimental results and the numerical ones.

\section{Discussion and Conclusion}\label{sec:Discussion_conclusion}

We have shown the pattern of the superposed state $\left| {\psi \left( 0 \right)} \right\rangle$ after scattered by a
single standing wave pulse. The larger the differences among the atomic numbers of peaks are, the lower the contrast is. However, by increasing the number of pulses, for the case of the same number of peaks, atomic densities for different peaks tend to be the same, which improves the contrast.

In sum, We developed a method of preparing the superposition of Bloch states in S-band with quasi-momenta spreading in FBZ, which is produced by the atomic oscillations in D-band when BECs confined in the superposed Bloch state of S and D bands of an OL at zero quasi-momentum evolve for a long time in the lattice and a harmonic trap. By acting the designed standing wave OL pulse on the superposed state, we can obtain certain patterns in the momentum space, with the intervals between the adjacent peaks far less than the double recoil momentum, where the minimum one we have given is $0.6 \hbar k_L$. The narrow interval is the result of the interference among several lowest bands corresponding to a single quasi-momentum under the action of the designed pulse, and the superposition of interference results with quasi-momenta over the FBZ. The experimental results are in good agreement with the theoretical ones.

\section{Acknowledgement}

We thank Guangjiong Dong, Peng Zheng, Biao Wu, Hongwei Xiong and Xuguang Yue for helpful discussion. This work is supported by the NKRDP (National Key Research and Development Program) under Grants No. 2016YFA0301500, NSFC (Grants No. 61475007, No. 11334001 and No. 91336103), and RFDP (Grants No. 20120001110091).

\bigskip
\emph{}

\end{document}